\documentclass{aa501}
\usepackage{psfig}
\usepackage{amsmath}
\usepackage{amssymb}

\newcommand{\MC}{\multicolumn}
\newcommand{\kms}{km\,s$^{-1}$}
\newcounter{qub}
\newcommand{\qq}{\addtocounter{qub}{1}\arabic{qub}}

\begin{document}

\title{ The Hamburg/SAO Survey for Emission--Line Galaxies }
\subtitle{ V. The Fifth List of 161 Galaxies }

\author{%
A.V.~Ugryumov\inst{1,11}
\and D.~Engels\inst{2}
\and A.Y.~Kniazev\inst{1,11}
\and R.F.~Green\inst{3}
\and Y.I.~Izotov\inst{4}\fnmsep\inst{9}
\and U.~Hopp\inst{5}\fnmsep\inst{10}
\and S.A.~Pustilnik\inst{1,11}
\and A.G.~Pramsky\inst{1,11}
\and T.F.~Kniazeva\inst{1}
\and N.~Brosch\inst{6}
\and H.-J.~Hagen\inst{2}
\and V.A.~Lipovetsky\inst{1}\fnmsep\thanks{Deceased 1996 September 22.}
\and J.~Masegosa\inst{7}
\and I.~M\'arquez\inst{7}
\and J.-M.~Martin\inst{8}
}

\offprints{A.Ugryumov  \email{and@sao.ru}}

\institute{
Special Astrophysical Observatory, Nizhnij Arkhyz, Karachai-Circassia,
369167, Russia
\and Hamburger Sternwarte, Gojenbergsweg 112, D-21029 Hamburg, Germany
\and National Optical Astronomical Observatories, Tucson, USA
\and Main Astronomical Observatory, Golosiiv, Kyiv-22, 03680, Ukraine
\and Universit\"atssternwarte M\"unchen, Scheiner Str. 1, D-81679 M\"unchen, Germany
\and Wise Observatory, Tel-Aviv University, Tel-Aviv 69978, Israel
\and Instituto de Astrofisica de Andalucia, CSIC, Aptdo. 3004, 18080, Granada, Spain
\and D\'epartement de Radioastronomie ARPEGES, Observatoire de Paris, F-92195 Meudon Cedex, France
\and Visiting astronomer at Kitt Peak National Observatory, USA
\and Visiting astronomer at Calar Alto Observatory, Spain
\and Isaac Newton Institute of Chile, SAO Branch
}

\date{Received \hskip 2cm; Accepted}

\abstract{
We present the fifth list with results\thanks{Tables 3 to 7 are
only available in electronic form at the CDS via anonymous ftp to
cdsarc.u-strasbg.fr (130.79.128.5) or via
http://cdsweb.u-strasbg.fr/Abstract.html. Figures A1 to A17 will be
made available only in the electronic version of the journal.} of the
Hamburg/SAO Survey for Emission-Line Galaxies (HSS therein, SAO --
Special Astrophysical Observatory, Russia).
The list is a result of follow-up spectroscopy conducted
with the 2.2\,m CAHA and 4\,m Kitt Peak telescopes in 1999.
The data of this snap-shot spectroscopy survey confirmed
166 emission-line objects out of 209 observed candidates and allowed their
quantitative spectral classification and redshift determination.
We could classify 98 emission-line objects as BCG/H{\sc ii}
galaxies or probable BCGs, 5 -- as QSOs, 3 -- as Seyfert galaxies,
2 -- as super-associations in subluminous spiral galaxies.
25 low-excitation objects were classified as starburst nuclei (SBN),
24 as dwarf amorphous nuclei starburst galaxies (DANS) and 3 as
LINERs. Due to low signal-to-noise ratio we could not classify 6 ELGs.
Furthermore, for another 4 galaxies we did not detect any significant
emission lines. For 131 emission-line galaxies, the redshifts and/or
line intensities are determined for the first time.
For the remaining 30 previously known ELGs we give either improved data
of the line intensities or some independent measurements.
}

\titlerunning{The Hamburg/SAO Survey for ELGs. V}
\authorrunning{A.V. Ugryumov et al.}

\maketitle

% TABLE 1. Summary of the samples in details

\begin{table*}
\begin{center}
\caption[]{\label{summary} 
Summary of the samples observed and breakdown of the classifications after
follow-up spectroscopy
}
\begin{tabular}{llccccccc}
\hline\noalign{\smallskip}
\MC{2}{c}{Candidate Sample}         &  N  & BCG  & Other & QSO &  Galaxies   & Stars & Not        \\
		    &               &     & \&   & ELGs  &     &  without    &       & Classified \\
		    &               &     & BCG? & \& SA &     &  emission   &       &            \\
\hline\noalign{\smallskip}
First priority
		    & new           & 141 & 64   & 43    &  5  &  3          & 18    &  8         \\
		    & already known &  28 & 21   &  7    & --  & --          & --    & --         \\
		    & total         & 169 & 85   & 50    &  5  &  3          & 18    &  8         \\[0.10cm]
\hline
Random sample       & new           &  16 & 3    &  1    & --  &  1          &  2    &  9         \\
		    & already known &   1 & --   &  1    & --  & --          & --    & --         \\
APM selected sample
& new           &  22 & 10   & 10    & --  & --          &  1    &  1         \\             %[0.10cm]
		    & already known &   1 & --   &  1    & --  & --          & --    & --         \\[0.10cm]
Second priority     & total         &  40 & 13   & 13    & --  &  1          &  3    & 10         \\[0.10cm]
\noalign{\smallskip}\hline\noalign{\smallskip}
\MC{2}{l}{Objects presented in this paper}
			      & 209 & 98   & 63    & 5   &  4          & 21    & 18         \\
\noalign{\smallskip}\hline
\end{tabular}
\end{center}
\end{table*}

% TABLE 2. JOURNAL OF OBSERVATIONS

\begin{table*}
\begin{center}
\caption{\label{Tab2} Journal of observations}
\begin{tabular}{ccccccc} \\ \hline
\MC{1}{c}{ Date } &
\MC{1}{c}{ Telescope }  &
\MC{1}{c}{ Instrument } &
\MC{1}{c}{ Grating,  }  &
\MC{1}{c}{ Wavelength } &
\MC{1}{c}{ Dispersion } &
\MC{1}{c}{ Observed } \\

\MC{1}{c}{ } & & &
\MC{1}{c}{ grism } &
\MC{1}{c}{ range [\AA] } &
\MC{1}{c}{ [\AA/pixel] } &
\MC{1}{c}{ number } \\

\MC{1}{c}{ (1) } &
\MC{1}{c}{ (2) } &
\MC{1}{c}{ (3) } &
\MC{1}{c}{ (4) } &
\MC{1}{c}{ (5) } &
\MC{1}{c}{ (6) } &
\MC{1}{c}{ (7) } \\
\hline
\\[-0.3cm]
09.06-18.06.1999 &          2.2~m CAHA & CAFOS   & G-200   & 3700--9500 & 4.5 & 117    \\
	     --- &                 --- &   ---   & B-200   & 3500--7400 & 4.7 & 18$^*$ \\
17.06-20.06.1999 & \MC{1}{r}{4~m KPNO} & R.C.Sp. & KPC-10A & 3700--8300 & 2.8 & 46     \\
08.12-11.12.1999 &          2.2~m CAHA & CAFOS   & G-200   & 3700--9500 & 9.0 & 46     \\
\hline \\[-0.2cm]
\MC{7}{l}{$^*$ -- objects reobserved
 }
\end{tabular}
\end{center}
\end{table*}

\section{Introduction}

The problem of creating large, homogeneous and deep samples of actively
star-forming low-mass galaxies is very important for several applications
in studies of galaxy evolution and spatial distribution.
Several earlier projects, like the Second Byurakan Survey (SBS) (Markarian et
al.~\cite{Markarian83}, Stepanian~\cite{Stepanian94}), the University of Michigan
(UM) survey (e.g., Salzer et al.~\cite{Salzer89}), and the Case survey (Pesch
et al.~\cite{Pesch95}, Salzer et al.~\cite{Salzer95}, Ugryumov et
al.~\cite{Ugryumov98}),
as well as some others, identified on objective prism plates many hundreds of
emission-line galaxies.
The Hamburg/SAO survey (HSS) is intended to create a new very large homogeneous
sample of such galaxies in the region of the Northern sky with an area
of some 1700 square degrees. The basic outline of the HSS and first results
are described in Paper I (Ugryumov et al.~\cite{Ugryumov99}), while
additional results from follow-up spectroscopy are given
in papers II, III and IV (Pustilnik et al. \cite{Pustilnik99},
Hopp et al. \cite{Hopp00}, Kniazev et al. \cite{Kniazev01}).
In this paper we present the results of follow-up spectroscopy of another 209
objects selected on the Hamburg Quasar Survey (HQS) prism spectral plates as
ELG candidates.

The article is organized as follows. In section \ref{Obs_red} we give the
details of the spectroscopic observations and of the data reduction.
In section \ref{Res_follow} the results of the observations are presented in
several tables.
In section \ref{Discussion} we
briefly discuss the new data and summarize the current state of the
Hamburg/SAO survey. Throughout this paper a Hubble constant H$_0$ = 75
km$\,$s$^{-1}$ Mpc$^{-1}$ is used.

\section{Spectral observations and data reduction}
\label{Obs_red}

\subsection{Observations}

The results presented here were obtained in a snap-shot observing
mode during two runs with the Calar Alto 2.2\,m and one run with the
KPNO 4\,m telescopes
(see Table~\ref{Tab2}).

\subsection{Observations with the KPNO 4\,m telescope}

The observations were carried out with the Ritchey-Chretien Spectrograph
attached to a Tektronix 2K$\times$2K~CCD detector. We used a
2$''$$\times$205$''$ slit with a KPC-10A grating (316 grooves~mm$^{-1}$)
in its first order, and a GG~375 order separation filter cutting
off second-order contamination for wavelengths blueward of
7400\,\AA. This instrumental setup allows a spatial scale along the slit of
$0\farcs69$~pixel$^{-1}$, a scale perpendicular to the slit of
2.77\,\AA\,pixel$^{-1}$, a spectral range of $3700-8300$\,\AA\
and a spectral resolution of $\sim$~7\,\AA\ (FWHM).
Short exposures ($3-5$ minutes) were used in order to detect strong emission
lines to allow measurement of redshifts and a crude classification.
No orientation of the slit along the parallactic angle was done because
of the snap-shot observing mode.
Reference spectra of an Ar-Ne-He lamp were recorded to provide
wavelength calibration.
Spectrophotometric standard stars from Oke (\cite{Oke90}) and Bohlin
(\cite{Bohlin96}) were observed at the beginning and at the end of each night
for flux calibration. The dome flats, bias, dark and twilight sky frames
were accumulated each night.
The weather conditions were photometric,
with seeing variations between 2\farcs5 and 3\arcsec\ (FWHM).

\subsection{Calar Alto 2.2\,m telescope observations}

Follow-up spectroscopy with the CAHA 2.2\,m telescope was carried out
during two runs (June and December, 1999, see Table~\ref{Tab2}), using
the Calar Alto Faint Object Spectrograph (CAFOS) and Cassegrain focal reducer.
During these runs a long slit of $300\arcsec \times (2\arcsec-3\arcsec)$ and
a G-200 grism (187\,\AA\,mm$^{-1}$, first order) were used.
The B-200 grism (185\,\AA\,mm$^{-1}$, first order) was also used to reobserve
18 objects in order to improve the [O{\sc ii}]\,$\lambda$\,3727\,\AA\ value.
There were no order separation filters applied.
The spatial scale along the slit was $0\farcs53$~pixel$^{-1}$.
A SITE~15 2K$\times$2K CCD was operated in a 2$\times$1 binned mode
during the December run (binning only along the dispersion direction),
while in the June run there was no binning applied.
The wavelength coverages were $\lambda$\,3700 -- $\lambda$\,9500\,\AA\
with maximum sensivity at $\sim$~6000\,\AA\
for the G-200 grism and $\lambda$\,3500 -- $\lambda$\,7400\,\AA\
with maximum sensivity at $\sim$~4000\,\AA\ for the B-200 grism.
The spectral resolution was
$\sim$~10\,\AA\ (FWHM) in the June run and due to CCD binning
$\sim$~20\,\AA\ (FWHM) in the December run.
The slit orientation was not aligned with the parallactic angle because of
the snap-shot observing mode. The exposure times varied within $2-15$ minutes
depending on the object brightness. The observations were complemented
by standard star flux measurements (Oke \cite{Oke90}, Bohlin \cite{Bohlin96}),
reference spectra (Hg-Cd lamp) for wavelength calibration, dome flat,
bias and dark frames.
Most of the time the weather conditions were photometric
with a seeing $\approx$1.5\arcsec\ (FWHM). Only during two nights
the weather conditions were variable with a seeing of
3\arcsec\ -- 4\arcsec (FWHM). The measurements of these nights are
marked by ``$*$'' in Table~4.

\subsection{Data reduction}

Reduction of the CAHA and KPNO spectral data was performed at the SAO
using the standard reduction systems MIDAS\footnote{ MIDAS is an acronym
for the European Southern Observatory package -- Munich Image Data Analysis
System.} and IRAF\footnote[2]{IRAF is distributed by
National Optical Astronomical Observatories, which is operated by the
Association of Universities for Research in Astronomy, Inc., under
cooperative agreement with the National Science Foundation}.

The MIDAS command FILTER/COSMIC was found to be a quite successful
way to remove automatically all cosmic ray hits from the images.
After that we applied the IRAF package CCDRED for bad pixel removal,
trimming, bias-dark subtraction, slit profile and flat-field corrections.

To do accurate wavelength calibration, correction for distortion
and tilt for each frame, sky substraction and correction for
atmospheric extinction, the IRAF package LONGSLIT was used with invoking
the IDENTIFY, REIDENTIFY, FITCOORD, TRANSFORM, BACKGROUND and EXTINCTION
tasks.

To obtain an instrumental response function from observed
spectrophotometric flux standards, the APSUM procedure from the APEXTRACT
package was used first to extract apertures of standard stars.
Then the sensivity curve determined by the STANDARD and SENSFUNC procedures
was applied by the CALIBRATE task to perform flux calibration for all
object images.
Finally the APSUM task was used to extract one-dimensional spectra
from the flux calibrated images.
In case that more than one exposure was obtained with the same setup
for an object, the extracted spectra were co-added and a mean vector
was calculated.
In case of several observations with different setups (telescopes or
grisms) for the same object, the data were reduced and measured
independently and the more reliable values were taken.

To speed-up and facilitate the line measurements we employed
dedicated command files created at the SAO using the FIT context and MIDAS
command language.
The procedures for the measurement of line parameters and redshifts
applied were described in detail in Papers III and IV.

\section{Results of follow--up spectroscopy}
\label{Res_follow}

In Table~\ref{summary} we present the results of the observations.
The 209 candidates were selected from our first and second priority
samples introduced in Paper IV.

Of 169 first priority candidates, 141 objects appeared in our list as
new ones. 28 objects were listed in the NED as galaxies or objects
from various catalogs with known redshifts and some of them already had
information on emission lines in earlier publications.
All objects were included in our observing program in order to improve spectral
information.
Comparison of our velocities with those of galaxies with already known 
redshift shows acceptable consistency within the uncertainties given.

Another 40 candidates observed were taken from the list of second priority
candidates. As described in Paper IV two samples were created from this list.
A ``random selected sample'' containing randomly selected objects from the list
and the ``APM selected sample'' which uses additional information for selection.
The ``random selected sample'' was created to access the fraction of BCGs in the
second priority list and contains 43 objects. For 26 of them the spectral
data were presented in Paper~IV, while for the remaining 17 candidates
observed with the CAHA 2.2\,m telescope, the spectral information is presented
here. The results of the analysis for this sample were presented in Paper~IV.
We found that the second priority list contains at most 10\% 
BCG/H{\sc ii}-galaxies.

The second, ``APM selected'' sample comprises second priority candidates which
are classified as non-stellar on Palomar Sky Survey plates (PSS) in the APM
database, and have blue color according to the APM color system
($(B-R) <$ 1.0).
Here we give spectral data for 23 objects from this sample. Except one
all are ELGs confirming the efficiency of this selection criterium to
pick up the BCG/H{\sc ii}-galaxies from the second priority list 
(cf. Paper~IV).

\subsection{Emission-line galaxies}

The observed emission line galaxies are listed in Table~3 containing
the following information: \\
 {\it column 1:} The object's IAU-type name with the prefix HS. \\
 {\it column 2:} Right ascension for equinox B1950. \\
 {\it column 3:} Declination for equinox B1950.
The coordinates were measured on direct plates of the HQS
and are accurate to $\sim$ 2$\arcsec$ (Hagen et al. \cite{Hagen95}). \\
 {\it column 4:} Heliocentric velocity and its r.m.s. uncertainty in
km~s$^{-1}$. \\
 {\it column 5:} Apparent B-magnitude obtained by calibration of the digitized
photoplates with photometric standard stars (Engels et al. \cite{Engels94}),
having an r.m.s. accuracy of $\sim$ $0\fm5$ for objects fainter than
m$_{\rm B}$ = $16\fm0$ (Popescu et al. \cite{Popescu96}).
Since the algorithm to calibrate the objective prism spectra is
optimized for point sources the brightnesses of extended galaxies are
underestimated. The resulting systematic uncertainties are expected to
be as large as 2 mag (Popescu et al. \cite{Popescu96}). For about 20\%
of our objects, B-magnitudes are unavailable at the moment. We present
for them blue magnitudes obtained from the APM database. They are
marked by a ``+'' before the value in the corresponding
column. According to our estimate they are systematically brighter by
$0\fm92$ than the B-magnitudes obtained by calibration of the
digitized photoplates (r.m.s.  $1\fm02$).
Objects referred to Popescu \& Hopp (\cite{Popescu00}) have
precise B-magnitude in Vennik, Hopp \& Popescu (\cite{Vennik2000}).
We do not list them here for the sake of homogeneity.
The B-magnitude for HS~1213+3636A was determined by eye estimate
as 17\fm5 and marked by a ``:'' as less confident. \\
 {\it column 6:} Absolute B-magnitude, calculated from the apparent
B-magnitude and the heliocentric velocity.
The only exception is made for HS~1213+3636B, a super-association in the
nearby galaxy NGC~4214 for which the distance is known from stellar
photometry (see comments below).
No correction for galactic extinction is made because all objects are
located at high galactic latitudes and the corrections are significantly
smaller than the uncertainties of the magnitudes. \\
 {\it column 7:} Preliminary spectral classification type according to
the spectral data presented in this article. BCG means a galaxy
possessing a characteristic H{\sc ii}-region spectrum with low enough
luminosity (M$_B \geq -$20$^m$). SBN and DANS are galaxies of lower
excitation with a corresponding position in line ratio diagrams, as
discussed in Paper~I. SBN are the brighter fraction of this type. Here we
follow the notation of Salzer et al. (\cite{Salzer89}).
Three objects were recognized as Seyfert galaxies.
Two of them (HS~1317+4521B and HS~1616+3627) are Sy1 galaxies due to
the presence of broad Balmer lines and of broad [Fe{\sc ii}] emission.
Our spectrum of HS~1616+3627 has insufficient quality to show this, but
independent spectroscopy data of Grupe et al.~(\cite{Grupe99}) clearly
classify this object as a narrow-line Sy1 galaxy.
The third one (HS~1220+3845) is a narrow-line ELG, which is classified as
a Sy2 galaxy in diagnostic diagrams.
Typical spectra of
low-ionisation nuclear emission-line regions (LINERs) are identified
for 3 galaxies.
SA stands for two probable super-associations in two dwarf spiral
galaxies. Six ELGs are difficult to classify. They are coded as NON. \\
 {\it column 8:} One or more alternative names, according to the
information from NED.
References are given to the other sources of the
redshift-spectral information indicating that a galaxy is an ELG.

The spectra of all emission-line galaxies are shown in Appendix~A,
which is available only in the electronic version of the journal.

The results of line flux measurements are given in Table~4.
It contains the following information: \\
 {\it column 1:} The object's IAU-type name with the prefix HS.
By asterisk we note the objects observed during non-photometric
conditions. \\
 {\it column 2:} Observed flux (in
10$^{-16}$\,erg\,s$^{-1}$\,cm$^{-2}$) of the H$\beta$\, line.
For several objects without H$\beta$ emission line the fluxes are given
for H$\alpha$ and marked by a ``+''.  \\
 {\it columns 3,4,5:} The observed flux ratios [O{\sc ii}]/H$\beta$,
[O{\sc iii}]/H$\beta$ and H$\alpha$/H$\beta$. \\
 {\it columns 6,7:} The observed flux ratios
[N{\sc ii}]\,$\lambda$\,6583\,\AA/H$\alpha$, and
([S{\sc ii}]\,$\lambda$\,6716\,\AA\ + \,$\lambda$\,6731\,\AA)/H$\alpha$. \\
 {\it columns 8,9,10:} Equivalent widths of the lines
[O{\sc ii}]\,$\lambda$\,3727\,\AA, H$\beta$ and
[O{\sc iii}]\,$\lambda$\,5007\,\AA.
For several objects without detected H$\beta$ emission line the equivalent
widths are given for H$\alpha$ and marked by a ``+''.
\\

\noindent
Below we give comments on some specific cases:
\\
\noindent
 {\it HS~0948+3723} and {\it HS~1430+4040}: H$\alpha$-emission in these
galaxies is affected by cosmic ray hits. The intensity ratio of
H$\gamma$/H$\beta$ is close to the recombination one with c$_{H\beta}=$0.
So we accepted for them the recombination H$\alpha$/H$\beta$ flux
ratio, and corrected respectively the ratios of [N{\sc ii}]/H$\alpha$ and
[S{\sc ii}]/H$\alpha$ in Table~4. \\
 {\it HS~1213+3636A}:
this is a compact slightly elongated ($\sim$9\arcsec) object at
$\sim$60\arcsec\ to  NW from the center of the star-bursting barred
Magellanic type galaxy (SBm) NGC~4214 (with M$_B=-$17\fm0).
Its radial velocity of 522$\pm$35 \kms\ is higher by $\sim$220 \kms\ than
the systemic velocity of NGC~4214 and higher by $\sim$260 \kms\ than
the velocity of H{\sc i} gas in this place (McIntyre~\cite{McIntyre97}).
This implies that this compact emission region is a kinematically detached
object, either a background dwarf, or more probably, a tiny ``satellite''
galaxy, passing by its much more massive neighbouring galaxy.
For the former case its distance from
the Hubble flow is 7~Mpc, M$_B=-$11\fm7: and the size $\sim$0.3~kpc.
For the latter case, if HS~1213+3636A is located at the same distance as
NGC~4214 (2.64 Mpc, as found through stellar photometry with HST by
Drozdovsky et al.~\cite{Drozd01}), its M$_B=-$9\fm6: and the size
$\sim$0.1 kpc. The nature of this tiny galaxy can be checked in
principle by the use of well known luminosity-abundance relationship for
gas-rich galaxies. If it is situated far from NGC~4214 one should
expect that this galaxy should have metallicity Z $\leq$ 1/30 Z$_{\odot}$.
On the other hand, if HS~1213+3636A is connected to NGC~4214, one can
speculate that this companion formed of gas pulled out of the parent SBm
galaxy
as a result of its strong interaction with other galaxy. In this case the
metallicity of this companion should be close to that of NGC~4214, that
is known to be $\sim$1/4--1/5 Z$_{\odot}$ (e.g., Kobulnitsky \& Skillman
\cite{KS96}).  Thus, metallicity determination of this dwarf with strong
starburst will probably resolve the dilemma.
\\
 {\it HS~1213+3636B}:
this well elongated object ($\sim$12\arcsec) is projected onto the middle part
($\sim$46\arcsec, or $\sim$0.6 kpc to NWW from the center) of the same galaxy
NGC~4214. It is at $\sim$25\arcsec\ to the South from HS~1213+3636A.
Our velocity differs by only $\sim$60 \kms\ from the NGC~4214 systemic velocity
and is well consistent with the H{\sc i} velocity of 260~\kms\ at this place from
McIntyre~(\cite{McIntyre97}). From our data it follows that HS~1213+3636B
is a super-association (SA) in a spiral arm of NGC~4214. Its M$_B=-$9\fm8:
and the size $\sim$0.15 kpc.
The spectra of both A and B objects were acquired in one slit, so their large
velocity difference is out of doubt. \\
 {\it HS~1311+3628}:
this elongated object ($\sim$11\arcsec) is projected onto the southern edge
($\sim$50\arcsec\ to the South from the center) of Im galaxy UGC~8303
(Holmberg~VIII).
A large difference of its radial velocity (1094$\pm$12~\kms) with that of
the systemic velocity of the dwarf irregular UGC~8303 (944$\pm$5 \kms\
from Huchra et al.~\cite{Huchra95}) and the local H{\sc i} velocity in this
region ($\sim$945 \kms) from the H{\sc i} map by Thean et al.~(\cite{Thean97})
(where this galaxy by error is called UGC~8314),
implies that this object probably does not belongs to UGC~8303 and is
a nearby separate star-forming dwarf.
We observed this object twice on different nights and with different
setups. Both spectra have the same redshifts within the observational
uncertainties.
HS~1311+3628 is therefore not an H{\sc ii}-region in UGC~8303 as
suggested by Popescu et al. (\cite{Popescu96}). \\
 {\it HS~1311+3924}:
this is the irregular (probably merging) galaxy UGC~8315 and member of a group,
with a radial velocity of 1215 \kms\ according to Garcia et al. (\cite{Garcia93}).
We got the spectrum of only one of two bright knots in the galaxy elongated
body, namely the NE one. While its velocity 1401$\pm$43 differs by 186 \kms\
from that given by Garcia et al. (\cite{Garcia93}) the combined uncertainty
of latter value is not well constrained. A long-slit spectrum along
the major axis will be helpful for understanding the nature of this object. \\
 {\it HS~1423+3945}: the radial velocity of this  object is very close to that
of UGC~9242 (our 1480$\pm$35 versus 1440$\pm$11~\kms\ in the RC3 catalog) and
therefore the object is very probably an SA in this edge-on dwarf spiral. \\
 {\it HS~1242+4058A}: The nearby BCG HS~1242+4058B (cf. Paper~II) is
separated by $\sim$4\arcmin\  from HS~1242+4058A and the radial velocities
differ by $\approx$300 km/s. They are probably not associated with each other. \\
 {\it HS~1317+4521B}: The nearby BCG-candidate ($\sim$10\arcsec\ to SW)
HS~1317+4521A is not observed yet. \\
 {\it HS~1646+4003B}: its faint companion ($\sim$30\arcsec\ to SSW),
HS~1646+4003A was earlier identified as NON (cf. Paper~II). \\
 {\it HS~0935+4135, HS~0958+3654, HS~1025+3707B, HS~1027+4728, HS~1206+4012,
HS~1236+4532, HS~1433+4245, HS~1514+4602}:
In all these galaxies the only emission line suitable for redshift measurement
is H$\alpha$.
A conservative estimate of their radial velocity uncertainty is
300~km\,s$^{-1}$. \\
For some of the galaxies either the snap-shot spectra had strong enough emission
lines and a well detected [O{\sc iii}]\,$\lambda$\,4363\,\AA\ or they were
reobserved later with better S/N ratio. For these galaxies we determined
metallicities and the following galaxies appeared to have
O/H~$\le$ 1/10~(O/H)$_{\odot}$:
HS~1313+4521, HS~1330+3651, HS~1334+3957, HS~1408+4227, HS~1417+4433,
HS~1545+4055, HS~1650+3706, HS~1655+3845. \\
For two BCGs (HS~0940+4052 and HS~1311+3628) the blue WR-bump is  more
or less evident from the snap-shot spectra, and for 8 more BCGs this
feature is probably present. \\
Several ``BCG?'' were listed in Table~3 with absolute magnitudes
brighter than $-20$\fm0. In fact, these M$_B$ are uncertain, since they come
from the estimates from the Hamburg Quasar Survey plates. Independent
CCD measurements or estimates from alternative sources show that usually they
are fainter by $0\fm5-1\fm0$, and should fulfill therefore the criterion
M$_B \geq -$20\fm0 (Kniazev et al.~\cite{Kn-HS37}).

\subsection{Quasars}

The main criteria applied to search for BCGs are a blue continuum near
$\lambda$\,4000\,\AA\  and a strong emission line --- expected is
[O{\sc iii}]\,$\lambda$\,5007\,\AA\ --- in
the wavelength region between 5000\,\AA\ and the sensitivity break of
the Kodak IIIa-J photoemulsion near 5400\,\AA\ (see Paper~I). For this
reason faint QSOs with Ly$\alpha$\,$\lambda$\,1216\,\AA\
redshifted to z $\sim$~3 or with Mg{\sc ii}\, $\lambda$\,2798\,\AA\
redshifted to z $\sim$~0.8 could be selected as
BCG candidates.
In Papers I--IV we already reported on the discovery of a
number of such faint QSOs. They have been missed by the proper Hamburg
Quasar Survey since the latter is restricted to bright
QSOs (B $\leq 17-17.5$). Here we report
on the discovery of five high-redshift faint (B $> 17.5$) QSOs.
For all but one of them, we identified Ly$\alpha$\,$\lambda$\,1216\,\AA\
redshifted to z $\sim$~3 as the line responsible for its selection. One
object (HS~1301+4233) shows a broad emission line tentatively identified with
Mg{\sc ii}\,$\lambda$\,2798\,\AA.
However this identification remains somewhat uncertain due to low
signal-to-noise of the spectrum.
Two of our new QSOs: HS~1224+4410 and HS~1541+4452, have recently appeared
in the NED as radio sources with no redshift determination (reported as
[HVG99]~R17 in Hunter et al.~(\cite{Hunter99}) and B3~1541+448A
in Douglas et al.~(\cite{Douglas96})).
The data for all five quasars are presented in
Table~5.  Finding charts and plots of their spectra can be found
on the www-site of the Hamburg Quasar Survey
(http://www.hs.uni-hamburg.de/hqs.html).

\subsection{Non-emission-line objects}

In total, for 43 candidates no (trustworthy) emission lines were detected.
We divided them into three categories.

\subsubsection{Absorption-line galaxies}

For four non-ELG galaxies the signal-to-noise ratio
of our spectra was sufficient to detect absorption lines, allowing
the determination of redshifts. The data are presented in Table~6.

\subsubsection{Stellar objects}

To separate the stars among the objects missing detectable emission
lines we cross-correlated a list of the most common stellar features
with the observed spectra.  In total, 21 objects with definite stellar
spectra and redshifts close to zero were identified.
All of them were crudely classified in categories
from definite A-stars to G-stars, with most of them intermediate
between A and F. The data for these stars are presented in Table~7.

\subsubsection{Non-classified  objects}

There was no possibility of classifying 18 objects without emission lines.
Their spectra have too low signal-to-noise ratio to detect trustworthy
absorption features, or the EWs of their emission lines are too small.

\section{Discussion}
\label{Discussion}

\subsection{The fifth list}

As result we have 209 observed candidates preselected on HQS objective
prism plates, of which 169 were first priority candidates
and 40 were second priority ones.
166 objects (79~\% from 209 objects) are found to be either ELGs (161),
or quasars (5).

Of 161 ELGs 98 galaxies (61~\%) were classified based on the character of
their  spectra and their luminosity as H{\sc ii}/BCGs or probable BCGs.

Two of them (HS~1213+3636A and HS~1311+3628)
are low-mass neighbours/satellites of the dwarf spiral NGC~4214 and of the
Im galaxy UGC~8303. As the discussion of the local environment of BCGs
is out of the scope of this paper, we refer to a recent analysis of this
issue by Pustilnik et al.~(\cite{Pustilnik01}).

HS~1213+3636B and HS~1423+3945 are  SAs in the dwarf spirals NGC~4214
and UGC~9242.

Six ELGs are difficult to classify at all due to their poor signal-to-noise
spectra. Six more ELGs were classified as Active Galactic Nuclei (AGNs):
3 as Seyfert galaxies and 3 as LINERs.
The remaining 49 ELGs are objects with low excitation:
either starburst nuclei galaxies (SBN and probable SBN) or their lower mass
analogs -- dwarf amorphous nuclear starburst galaxies (DANS or probable DANS).

By keeping a high fraction of BCGs ($\sim$ 62~\% among the first priority
candidates) we continue to have a high efficiency of discovery new BCGs,
which is the main goal of the HSS.
Since the completeness of the BCG sample under construction is an important
parameter for many follow-up statistical studies, we observed a randomly
selected sample of candidates from our list of second priority candidates.
As discussed already in Paper~IV at most 10~\% of them turned out to be BCGs.
Using additional information from the APM is an efficient means to pick up
these BCGs among the second priority candidates: among 23 objects observed
10 BCGs (43~\%) were discovered (Table~\ref{summary}).

\subsection{Summary of the present status of the survey}

Summarizing the results of the Hamburg/SAO survey presented in Papers~I
through V, we discovered altogether from the 1-st priority candidates
433 new emission-line objects (25 of them are QSOs), and for 85 known ELGs
we got quantitative data for their emission lines.
At the moment the total number of confident or probable blue compact/low-mass
H{\sc ii}-galaxies reaches 360. Relative to all observed 493 ELGs
the fraction of BCGs (360/493 or 73~\%) demonstrates the high efficiency of
the survey to find this type of galaxies.
21 more new BCGs and 20 other type ELGs are found among the second priority
candidates.
To estimate the total number of BCGs in the HSS zone we should count
new BCGs expected from the remaining candidates and those selected
in the HSS, but not observed by us since they already were known from
other surveys. Thus we expect the total number of BCGs in this sky region
to be $\sim$500.
This will be the largest homogeneous BCG sample in both hemispheres.

\section{Conclusions}

We made follow-up spectroscopy of the fifth list of candidates from the
Hamburg/SAO Survey for ELGs.
Summarizing the results of the spectroscopy, the analysis of spectral
information and the discussion above we draw the following conclusions:

\begin{itemize}

\item The intended methods to detect ELG candidates on the plates of the
      Hamburg Quasar Survey give a reasonably high detection rate of
      emission-line objects. In total, out of both priority categories,
      we observed 209 objects among which we found 166 emission-line
      objects corresponding to a detection rate of $\sim$~79~\%.

\item Besides the ELGs we found also 5 new quasars, mostly with Ly$\alpha$
      in the wavelength region $4950-5100$\,\AA\ (z~$\sim$~3),
      near the red boundary of the IIIa-J photoplates.

\item The high fraction of BCG/H{\sc ii} galaxies among all observed
      ELGs (about 61~\% in this paper) is in line with our main
      goal -- to pick up efficiently a statistically well selected and
      deep BCG sample in the sky region under analysis.

\end{itemize}

\begin{acknowledgements}

This work was supported by the grant of the Deutsche
Forschungsgemeinschaft No.~436 RUS~17/77/94. U.A.V. is very grateful
to the staff of the Hamburg Observatory for their hospitality and kind
assistance.  Support by the INTAS grant No.~96-0500 was crucial to
proceed with the Hamburg/SAO survey declination band centered on
+37.5$^{\circ}$. SAO authors acknowledge also partial support from
the INTAS grant No.~97-0033.
We note that the use of APM facility was extremely valuable
for selection methodology of additional candidates to BCGs from the 2-nd
priority list.
This research has made use of the
NASA/IPAC Extragalactic Database (NED) which is operated by the Jet
Propulsion Laboratory, California Institute of Technology, under contract
with the National Aeronautics and Space Administration. We have
also used the Digitized Sky Survey, produced at the Space Telescope Science
Institute under government grant NAG W-2166.

\end{acknowledgements}

\clearpage
\renewcommand{\baselinestretch}{1.3}

%\begin{document}

\scriptsize
%\newcounter{qub}
\setcounter{qub}{0}

\begin{table*}[h]

\begin{center}
\caption{\label{Tab3} Coordinates, Velocities and Magnitudes of
Emission--Line Galaxies}

\begin{tabular}{rlllrrcll} \hline \\[-0.35cm]
\multicolumn{1}{c}{\#}               &
\multicolumn{1}{c}{Name}             &
\multicolumn{1}{c}{$\alpha\,(1950)$} &
\multicolumn{1}{c}{$\delta\,(1950)$} &
\multicolumn{1}{c}{$V_{hel}^{\,a}\pm\sigma_V$}    &
\multicolumn{1}{r}{m$_{B}$}          &
\multicolumn{1}{c}{M$_{B}^{\;\;b}$}  &
\multicolumn{1}{c}{Type}             &
\multicolumn{1}{l}{Other names from NED}\\
&&&&&&&& \multicolumn{1}{l}{and number of reference}\\
&
\multicolumn{1}{c}{ (1) } &
\multicolumn{1}{c}{ (2) } &
\multicolumn{1}{c}{ (3) } &
\multicolumn{1}{c}{ (4) } &
\multicolumn{1}{c}{ (5) } &
\multicolumn{1}{c}{ (6) } &
\multicolumn{1}{c}{ (7) } &
\multicolumn{1}{c}{ (8) } \\
\hline
%-------------------------------------------------------------------------------------------------------------
% # |   IAU_name             |RA_1950    |DEC_1950  |V0      +-  V_err|mag   |Mb     |Type      |Other Names           |
%-------------------------------------------------------------------------------------------------------------
\qq& HS 0809+4334            &08 09 01.5 &+43 34 17 &14146 $\pm$ 57   & 18.4 &--18.0 & BCG      &  \\
\qq& HS 0818+4234            &08 18 40.5 &+42 34 23 &17424 $\pm$ 68   & 19.7 &--17.1 & NON      &  \\
\qq& HS 0822+4049            &08 22 25.1 &+40 49 57 &12087 $\pm$ 100  & 17.3 &--18.7 & BCG      &  \\
\qq& HS 0822+4238            &08 22 50.9 &+42 38 24 &11637 $\pm$ 60   & 18.2 &--17.8 & BCG      &  \\
\hline
\end{tabular}
\end{center}
\end{table*}

\newpage

\begin{table*}[h]

\begin{center}
\flushleft {\bf Table 3.} (Continued)

\begin{tabular}{rlllrrcll} \\[-0.22cm] \hline \\[-0.35cm]
\multicolumn{1}{c}{\#}               &
\multicolumn{1}{c}{Name}             &
\multicolumn{1}{c}{$\alpha\,(1950)$} &
\multicolumn{1}{c}{$\delta\,(1950)$} &
\multicolumn{1}{c}{$V_{hel}^{\,a}\pm\sigma_V$}    &
\multicolumn{1}{r}{m$_{B}$}          &
\multicolumn{1}{c}{M$_{B}^{\;\;b}$}  &
\multicolumn{1}{c}{Type}             &
\multicolumn{1}{l}{Other names from NED}\\
&&&&&&&& \multicolumn{1}{l}{and number of reference}\\
&
\multicolumn{1}{c}{ (1) } &
\multicolumn{1}{c}{ (2) } &
\multicolumn{1}{c}{ (3) } &
\multicolumn{1}{c}{ (4) } &
\multicolumn{1}{c}{ (5) } &
\multicolumn{1}{c}{ (6) } &
\multicolumn{1}{c}{ (7) } &
\multicolumn{1}{c}{ (8) } \\
\hline
%-------------------------------------------------------------------------------------------------------------
% # |   IAU_name             |RA_1950    |DEC_1950  |V0      +-  V_err|mag   |Mb     |Type      |Other Names           |
%-------------------------------------------------------------------------------------------------------------
\qq& HS 0958+3718            &09 58 15.5 &+37 18 38 &1606  $\pm$ 77   & 15.7 &--16.0 & BCG      & CG~315,KUG0958+373,{\bf 5} \\
\qq& HS 1004+3730$^\ddagger$ &10 04 22.3 &+37 30 22 &10133 $\pm$ 94   & 16.5 &--19.2 & SBN      & KUG1004+375 \\
\qq& HS 1008+3620$^\ddagger$ &10 08 01.5 &+36 20 51 &11424 $\pm$ 53   & 17.2 &--18.7 & BCG      &  \\
\hline
\end{tabular}
\end{center}
\end{table*}

\newpage

\begin{table*}[h]

\begin{center}
\flushleft {\bf Table 3.} (Continued)

\begin{tabular}{rlllrrcll} \\[-0.22cm] \hline \\[-0.35cm]
\multicolumn{1}{c}{\#}               &
\multicolumn{1}{c}{Name}             &
\multicolumn{1}{c}{$\alpha\,(1950)$} &
\multicolumn{1}{c}{$\delta\,(1950)$} &
\multicolumn{1}{c}{$V_{hel}^{\,a}\pm\sigma_V$}    &
\multicolumn{1}{r}{m$_{B}$}          &
\multicolumn{1}{c}{M$_{B}^{\;\;b}$}  &
\multicolumn{1}{c}{Type}             &
\multicolumn{1}{l}{Other names from NED}\\
&&&&&&&& \multicolumn{1}{l}{and number of reference}\\
&
\multicolumn{1}{c}{ (1) } &
\multicolumn{1}{c}{ (2) } &
\multicolumn{1}{c}{ (3) } &
\multicolumn{1}{c}{ (4) } &
\multicolumn{1}{c}{ (5) } &
\multicolumn{1}{c}{ (6) } &
\multicolumn{1}{c}{ (7) } &
\multicolumn{1}{c}{ (8) } \\
\hline
%-------------------------------------------------------------------------------------------------------------
% # |   IAU_name             |RA_1950    |DEC_1950  |V0      +-  V_err|mag   |Mb     |Type      |Other Names           |
%-------------------------------------------------------------------------------------------------------------
\qq& HS 1236+4532            &12 36 39.8 &+45 32 54 &16334:$\pm$ 300  & 17.6 &--19.1 & NON      & MAPS-NGPO 172 027192 \\
\qq& HS 1237+3900            &12 37 04.0 &+39 00 00 &11198 $\pm$ 11   &+18.6 &--17.3 & BCG      & {\bf 6} \\
\qq& HS 1238+4229            &12 38 24.1 &+42 29 08 &11370 $\pm$ 53   & 18.9 &--17.0 & BCG      &  \\
\hline
\end{tabular}
\end{center}
\end{table*}

\newpage

\begin{table*}[h]

\begin{center}
\flushleft {\bf Table 3.} (Continued)

\begin{tabular}{rlllrrcll} \\[-0.22cm] \hline \\[-0.35cm]
\multicolumn{1}{c}{\#}               &
\multicolumn{1}{c}{Name}             &
\multicolumn{1}{c}{$\alpha\,(1950)$} &
\multicolumn{1}{c}{$\delta\,(1950)$} &
\multicolumn{1}{c}{$V_{hel}^{\,a}\pm\sigma_V$}    &
\multicolumn{1}{r}{m$_{B}$}          &
\multicolumn{1}{c}{M$_{B}^{\;\;b}$}  &
\multicolumn{1}{c}{Type}             &
\multicolumn{1}{l}{Other names from NED}\\
&&&&&&&& \multicolumn{1}{l}{and number of reference}\\
&
\multicolumn{1}{c}{ (1) } &
\multicolumn{1}{c}{ (2) } &
\multicolumn{1}{c}{ (3) } &
\multicolumn{1}{c}{ (4) } &
\multicolumn{1}{c}{ (5) } &
\multicolumn{1}{c}{ (6) } &
\multicolumn{1}{c}{ (7) } &
\multicolumn{1}{c}{ (8) } \\
\hline
%-------------------------------------------------------------------------------------------------------------
% # |   IAU_name             |RA_1950    |DEC_1950  |V0      +-  V_err|mag   |Mb     |Type      |Other Names           |
%-------------------------------------------------------------------------------------------------------------
\qq& HS 1419+3639            &14 19 16.2 &+36 39 18 &3579  $\pm$ 42   & 17.1 &--16.3 & BCG      & UGC~9198,{\bf 2} \\
\qq& HS 1420+3726$^\dagger$  &14 20 50.1 &+37 26 58 &10847 $\pm$ 45   & 17.5 &--18.3 & BCG      &  \\
\qq& HS 1423+3945$^\dagger$  &14 23 13.6 &+39 45 27 &1480  $\pm$ 35   & 18.5 &--13.0 & SA in Sp & in UGC~9242 \\
\end{tabular}
\end{center}
\end{table*}
%\end{document}

\newpage

\begin{table*}[h]

\begin{center}
\flushleft {\bf Table 3.} (Continued)
\begin{tabular}{rlllrrcll} \\[-0.22cm]
%\hline \\[-0.35cm]
\multicolumn{9}{l}{{\bf References:}} \\
\multicolumn{9}{l}{{\bf 1 } -- Klemola, A. R., Jones, B. F., \& Hanson, R. B.  1987, AJ, 94, 501                                                 } \\
\multicolumn{9}{l}{{\bf 2 } -- Popescu, C. C., Hopp, U., Hagen, H.-J., \& Els\"{a}sser, H.  1996, A\&AS, 116, 43                                 } \\
\multicolumn{9}{l}{{\bf 3 } -- Ugryumov, A. V., Pustilnik, S. A., Lipovetsky, V. A., Izotov, Yu. I., \& Richter, G. M.  1998, A\&AS, 131, 295    } \\
\multicolumn{9}{l}{{\bf 4 } -- RC3                                                                                                               } \\
\multicolumn{9}{l}{{\bf 5 } -- Weistrop, D., \& Downes, R. A.  1991, AJ, 102, 1680                                                               } \\
\multicolumn{9}{l}{{\bf 6 } -- Hopp, U., Engels, D., Green, R., et al.  2000, A\&AS, 142, 417 ({\bf Paper~III})                                  } \\
\end{tabular}
\end{center}
\end{table*}

%\end{document}

%\documentstyle[psfig]{aa}
%\pagestyle{empty}
\renewcommand{\baselinestretch}{1.4}

%\begin{document}

\scriptsize
%\newcounter{qub}
\setcounter{qub}{0}

\begin{table*}[h]

\begin{center}
\caption{\label{Tab4} {\bf Emission line parameters}}

\begin{tabular}{rlrcccccrrr} \hline \\[-0.35cm]
\multicolumn{1}{c}{\#}                             &
\multicolumn{1}{c}{ Name}                          &
\multicolumn{1}{c}{ F(H$\beta$)$^a$ }              &
\multicolumn{1}{c}{ $\frac{F(\lambda 3727)}{F(H\beta)}$} &
\multicolumn{1}{c}{ $\frac{F(\lambda 5007)}{F(H\beta)}$} &
\multicolumn{1}{c}{ $\frac{F(H\alpha)}{F(H\beta)}$} &
\multicolumn{1}{c}{ $\frac{F(\lambda 6583)}{F(H\alpha)}$}&
\multicolumn{1}{c}{ $\frac{F([SII])}{F(H\alpha)}$} &
\multicolumn{1}{c}{ W$_{\lambda 3727}$(\AA)}       &
\multicolumn{1}{c}{ W$_{H\beta}$(\AA)}             &
\multicolumn{1}{c}{ W$_{\lambda 5007}$(\AA)}       \\
&
\multicolumn{1}{c}{ (1) } &
\multicolumn{1}{c}{ (2) } &
\multicolumn{1}{c}{ (3) } &
\multicolumn{1}{c}{ (4) } &
\multicolumn{1}{c}{ (5) } &
\multicolumn{1}{c}{ (6) } &
\multicolumn{1}{c}{ (7) } &
\multicolumn{1}{c}{ (8) } &
\multicolumn{1}{c}{ (9) } &
\multicolumn{1}{c}{ (10) } \\
\hline
%-------------------------------------------------------------------------------------------------------
%    IAU_name                FL_4861 F3727_Hb F5007_Hb Ha_Hb  F6583_Ha FSii_Ha EW_3727 EW_4861 EW_5007
%-------------------------------------------------------------------------------------------------------
\qq& HS 0809+4334            &51     &1.16    &3.96    &3.71  &0.16    &0.20   &41     &24     &92    \\
\qq& HS 0818+4234            &+5:    &        &        &      &        &0.75:  &       &+12:   &5:    \\
\qq& HS 0822+4049            &85     &1.45    &2.25    &4.62  &0.14    &0.20   &72     &38     &86    \\
\qq& HS 0822+4238            &16     &2.81    &3.94    &4.63  &0.08    &0.16   &39     &14     &55    \\
\qq& HS 0824+4139            &11     &        &3.55    &4.64  &0.08    &0.41   &       &8      &31    \\
\end{tabular}
\end{center}
\end{table*}

\newpage

\begin{table*}[h]

\begin{center}
\flushleft {\bf Table 4.} (Continued)
\begin{tabular}{rlrcccccrrr} \hline \\[-0.35cm]
\multicolumn{1}{c}{\#}                             &
\multicolumn{1}{c}{ Name}                          &
\multicolumn{1}{c}{ F(H$\beta$)$^a$ }              &
\multicolumn{1}{c}{ $\frac{F(\lambda 3727)}{F(H\beta)}$} &
\multicolumn{1}{c}{ $\frac{F(\lambda 5007)}{F(H\beta)}$} &
\multicolumn{1}{c}{ $\frac{F(H\alpha)}{F(H\beta)}$} &
\multicolumn{1}{c}{ $\frac{F(\lambda 6583)}{F(H\alpha)}$}&
\multicolumn{1}{c}{ $\frac{F([SII])}{F(H\alpha)}$} &
\multicolumn{1}{c}{ W$_{\lambda 3727}$(\AA)}       &
\multicolumn{1}{c}{ W$_{H\beta}$(\AA)}             &
\multicolumn{1}{c}{ W$_{\lambda 5007}$(\AA)}       \\
&
\multicolumn{1}{c}{ (1) } &
\multicolumn{1}{c}{ (2) } &
\multicolumn{1}{c}{ (3) } &
\multicolumn{1}{c}{ (4) } &
\multicolumn{1}{c}{ (5) } &
\multicolumn{1}{c}{ (6) } &
\multicolumn{1}{c}{ (7) } &
\multicolumn{1}{c}{ (8) } &
\multicolumn{1}{c}{ (9) } &
\multicolumn{1}{c}{ (10) } \\
\hline
%-------------------------------------------------------------------------------------------------------
%    IAU_name                FL_4861 F3727_Hb F5007_Hb Ha_Hb  F6583_Ha FSii_Ha EW_3727 EW_4861 EW_5007
%-------------------------------------------------------------------------------------------------------
\qq& HS 0958+3718$^*$        &190    &0.24    &6.46    &2.25  &0.03    &       &21     &64     &430   \\
\qq& HS 1004+3730$^\ddagger$ &+81    &        &        &      &0.42    &0.40   &       &+38    &3     \\
\qq& HS 1008+3620$^\ddagger$ &13     &        &6.15    &6.31  &0.07    &0.22   &       &10     &64    \\
\qq& HS 1011+3604            &26     &1.81    &2.08    &3.85  &0.16    &0.29   &97     &15     &33    \\
\qq& HS 1014+3535$^\ddagger$ &5:     &        &        &16:   &0.37    &0.25   &       &2:     &      \\
\end{tabular}
\end{center}
\end{table*}

\newpage

\begin{table*}[h]

\begin{center}
\flushleft {\bf Table 4.} (Continued)
\begin{tabular}{rlrcccccrrr} \hline \\[-0.35cm]
\multicolumn{1}{c}{\#}                             &
\multicolumn{1}{c}{ Name}                          &
\multicolumn{1}{c}{ F(H$\beta$)$^a$ }              &
\multicolumn{1}{c}{ $\frac{F(\lambda 3727)}{F(H\beta)}$} &
\multicolumn{1}{c}{ $\frac{F(\lambda 5007)}{F(H\beta)}$} &
\multicolumn{1}{c}{ $\frac{F(H\alpha)}{F(H\beta)}$} &
\multicolumn{1}{c}{ $\frac{F(\lambda 6583)}{F(H\alpha)}$}&
\multicolumn{1}{c}{ $\frac{F([SII])}{F(H\alpha)}$} &
\multicolumn{1}{c}{ W$_{\lambda 3727}$(\AA)}       &
\multicolumn{1}{c}{ W$_{H\beta}$(\AA)}             &
\multicolumn{1}{c}{ W$_{\lambda 5007}$(\AA)}       \\
&
\multicolumn{1}{c}{ (1) } &
\multicolumn{1}{c}{ (2) } &
\multicolumn{1}{c}{ (3) } &
\multicolumn{1}{c}{ (4) } &
\multicolumn{1}{c}{ (5) } &
\multicolumn{1}{c}{ (6) } &
\multicolumn{1}{c}{ (7) } &
\multicolumn{1}{c}{ (8) } &
\multicolumn{1}{c}{ (9) } &
\multicolumn{1}{c}{ (10) } \\
\hline
%-------------------------------------------------------------------------------------------------------
% # |   IAU_name             FL_4861 F3727_Hb F5007_Hb Ha_Hb  F6583_Ha FSii_Ha EW_3727 EW_4861 EW_5007
%-------------------------------------------------------------------------------------------------------
\qq& HS 1236+4532            &+3:    &        &        &      &        &       &       &+3:    &      \\
\qq& HS 1237+3900            &40     &2.33    &5.35    &2.90  &0.07:   &0.11:  &79     &39     &227   \\
\qq& HS 1238+4229            &15     &5.07    &4.93    &4.53  &0.09    &0.25:  &86     &18     &91    \\
\qq& HS 1242+4058A$^*$       &47     &        &2.94    &4.09  &0.08    &0.27   &       &18     &53    \\
\qq& HS 1244+3647$^*$        &110    &0.89    &4.01    &2.55  &0.03    &0.32   &22     &37     &150   \\
\end{tabular}
\end{center}
\end{table*}

\newpage

\begin{table*}[h]

\begin{center}
\flushleft {\bf Table 4.} (Continued)
\begin{tabular}{rlrcccccrrr} \hline \\[-0.35cm]
\multicolumn{1}{c}{\#}                             &
\multicolumn{1}{c}{ Name}                          &
\multicolumn{1}{c}{ F(H$\beta$)$^a$ }              &
\multicolumn{1}{c}{ $\frac{F(\lambda 3727)}{F(H\beta)}$} &
\multicolumn{1}{c}{ $\frac{F(\lambda 5007)}{F(H\beta)}$} &
\multicolumn{1}{c}{ $\frac{F(H\alpha)}{F(H\beta)}$} &
\multicolumn{1}{c}{ $\frac{F(\lambda 6583)}{F(H\alpha)}$}&
\multicolumn{1}{c}{ $\frac{F([SII])}{F(H\alpha)}$} &
\multicolumn{1}{c}{ W$_{\lambda 3727}$(\AA)}       &
\multicolumn{1}{c}{ W$_{H\beta}$(\AA)}             &
\multicolumn{1}{c}{ W$_{\lambda 5007}$(\AA)}       \\
&
\multicolumn{1}{c}{ (1) } &
\multicolumn{1}{c}{ (2) } &
\multicolumn{1}{c}{ (3) } &
\multicolumn{1}{c}{ (4) } &
\multicolumn{1}{c}{ (5) } &
\multicolumn{1}{c}{ (6) } &
\multicolumn{1}{c}{ (7) } &
\multicolumn{1}{c}{ (8) } &
\multicolumn{1}{c}{ (9) } &
\multicolumn{1}{c}{ (10) } \\
\hline
%-------------------------------------------------------------------------------------------------------
%    IAU_name                FL_4861 F3727_Hb F5007_Hb Ha_Hb  F6583_Ha FSii_Ha EW_3727 EW_4861 EW_5007
%-------------------------------------------------------------------------------------------------------
\qq& HS 1419+3639            &69     &3.74    &2.43    &1.96  &0.13    &0.31   &186    &41     &102   \\
\qq& HS 1420+3726$^{* \dagger}$&34     &2.53    &3.21    &2.56  &        &0.34   &69     &35     &120   \\
\qq& HS 1423+3945$^\dagger$  &70     &        &4.56    &2.24  &0.03    &0.11   &       &136    &638   \\
\qq& HS 1423+4054            &25     &4.12    &2.72    &1.96  &0.08    &0.20   &64     &18     &54    \\
\qq& HS 1427+4309$^*$        &34     &2.24    &4.09    &2.91  &0.08    &0.26   &54     &23     &99    \\
\end{tabular}
\end{center}
\end{table*}

%\end{document}

%\documentstyle[psfig]{aa}
%\pagestyle{empty}
\renewcommand{\baselinestretch}{1.4}

%\begin{document}

\setcounter{qub}{0}

\begin{table*}[h]

\begin{center}
\caption{\label{Tab5} Coordinates, Redshifts and Magnitudes of QSOs}

\begin{tabular}{rlcclrl} \hline \\[-0.35cm]
\MC{1}{c}{\#}               &
\MC{1}{c}{Name}             &
\MC{1}{c}{$\alpha\,(1950)$} &
\MC{1}{c}{$\delta\,(1950)$} &
\MC{1}{c}{$z^{\,a}$}        &
\MC{1}{c}{$m_{B}$}          &
\MC{1}{c}{Detected {\bf emission lines} }\\
&
\MC{1}{c}{ (1) } &
\MC{1}{c}{ (2) } &
\MC{1}{c}{ (3) } &
\MC{1}{c}{ (4) } &
\MC{1}{c}{ (5) } &
\MC{1}{c}{ (6) }
\\ \hline
%-------------------------------------------------------------------------------------
%IAU_name                   |RA_1950    |DEC_1950   |QZ_M   |mag   | Lines           |
%-------------------------------------------------------------------------------------
\qq & HS 0936+4204          &09 36 59.6 &+42 04 24  &3.3211 &18.3  & Ly$\alpha$ 1216 \AA, Si{\sc iv}/O{\sc iv}] 1400 \AA, C{\sc iv} 1549 \AA  \\
\qq & HS 1009+3706          &10 09 02.4 &+37 06 31  &3.0393 &17.6  & Ly$\alpha$ 1216 \AA, Si{\sc iv}/O{\sc iv}] 1400 \AA, C{\sc iv} 1549 \AA \\
\qq & HS 1224+4410          &12 24 55.1 &+44 10 03  &3.0718 &19.5  & Ly$\alpha$ 1216 \AA, C{\sc iv} 1549 \AA \\
\qq & HS 1301+4233          &13 01 47.0 &+42 33 30  &0.8460:&+18.7 & Mg{\sc ii} 2798 \AA \\
\qq & HS 1541+4452          &15 41 23.0 &+44 52 27  &2.3588 &18.8  & Ly$\alpha$ 1216 \AA, Si{\sc iv}/O{\sc iv}] 1400 \AA, C{\sc iv} 1549 \AA \\
\hline \\[-0.2cm]
\multicolumn{7}{l}{ $^a$ Observed redshift; $^+$ APM magnitudes; : less confident values.} \\
\end{tabular}
\end{center}
\end{table*}

%\end{document}

%\documentstyle[psfig]{l-aa}
%\pagestyle{empty}
\renewcommand{\baselinestretch}{1.4}

%\begin{document}

\setcounter{qub}{0}
\begin{table*}[h]

\begin{center}
\caption{\label{Tab6} Galaxies without detected emission lines }

\begin{tabular}{rllllccl} \hline \\[-0.35cm]
\multicolumn{1}{c}{\#}   &
\multicolumn{1}{c}{Name} &
\multicolumn{1}{c}{$\alpha\,(1950)$} &
\multicolumn{1}{c}{$\delta\,(1950)$} &
\multicolumn{1}{c}{$v^{\,a}_{0}$} &
\multicolumn{1}{c}{m$_B$} &
\multicolumn{1}{r}{M$_B^{\,b}$} &
\multicolumn{1}{c}{Absorption lines} \\
&
\MC{1}{c}{ (1) } &
\MC{1}{c}{ (2) } &
\MC{1}{c}{ (3) } &
\MC{1}{c}{ (4) } &
\MC{1}{c}{ (5) } &
\MC{1}{c}{ (6) } &
\MC{1}{c}{ (7) }
\\ \hline
\qq& HS 1008+3538           & 10 08 00.1 &+35 38 02  &13968:&17.4 &--19.0  & H$\beta$, MgIb, H$\alpha$ \\
\qq& HS 1207+3745           & 12 07 38.0 &+37 45 53  &13381:&18.8 &--17.5  & H$\beta$, H$\alpha$ \\
\qq& HS 1408+3524$^\dagger$ & 14 08 32.6 &+35 24 23  &19933 &16.8 &--20.3  & CaK, CaH, G$_{band}$, H$\beta$, MgIb, CaFe5269, NaD \\
\qq& HS 1419+4026           & 14 19 50.0 &+40 26 28  &14235:&18.1 &--18.3  & G$_{band}$, H$\alpha$ \\
\hline \\[-0.2cm]
\multicolumn{8}{l}{ $^a$ Heliocentric velocities; $^b$ absolute magnitudes are not corrected for galactic extinction;} \\
\multicolumn{8}{l}{ $^\dagger$ objects from the second priority random sample; : less confident values.} \\

\end{tabular}
\end{center}

\end{table*}

%\end{document}

%\documentstyle[psfig]{l-aa}
%\documentstyle[psfig]{article}
%\pagestyle{empty}

%\begin{document}

\setcounter{qub}{0}

\begin{table*}[t]
\begin{center}
\caption{\label{Tab7}Stars}
\begin{tabular}{rlccrcl} \hline \\[-0.35cm]
\multicolumn{1}{c}{\#}             &
\multicolumn{1}{c}{Name}             &
\multicolumn{1}{c}{$\alpha\,(1950)$} &
\multicolumn{1}{c}{$\delta\,(1950)$} &
\multicolumn{1}{r}{m$_{B}$}          &
\multicolumn{1}{l}{Type}             &
\multicolumn{1}{c}{Spectral features} \\
&
\MC{1}{c}{ (1) } &
\MC{1}{c}{ (2) } &
\MC{1}{c}{ (3) } &
\MC{1}{c}{ (4) } &
\MC{1}{c}{ (5) } &
\MC{1}{c}{ (6) }
\\ \hline
\qq &HS 0852+4257           & 08 52 23.3 & +42 57 22 &+20.2 & A  & H$_{\gamma}$, H$_{\beta}$, H$_{\alpha}$ \\
\qq &HS 0945+4851           & 09 45 00.5 & +48 51 00 &+18.0 &F-G & H$_{\beta}$, MgIb, CaFe5269, H$_{\alpha}$ \\
\qq &HS 1012+3815           & 10 12 45.4 & +38 15 01 & 17.2 & A  & H$_{\gamma}$, H$_{\beta}$, H$_{\alpha}$ \\
\qq &HS 1013+3804$^\dagger$ & 10 13 05.6 & +38 04 27 & 17.7 & F  & H$_{\beta}$, NaD, H$_{\alpha}$ \\
\qq &HS 1029+4020$^\dagger$ & 10 29 10.6 & +40 20 35 & 16.7 & G  & MgIb, CaFe5269, NaD, H$_{\alpha}$ \\
\end{tabular}
\end{center}
\end{table*}

%\end{document}

%\clearpage

% APPENDIX

%\input appendix


\begin{thebibliography}{99}

\bibitem[1996]{Bohlin96}
 Bohlin, R. C.
 1996, AJ, 111, 1743
\bibitem[1996]{Douglas96}
 Douglas, J. N., Bash, F. N., Bozyan, F. A., Torrence, G. W., \& Wolfe, C.
 1996, AJ, 111, 1945
\bibitem[2001]{Drozd01}
 Drozdovsky, I., et al.
 2001, in Bad Honnef conference ``Dwarf galaxies and their environment'',
 Jan. 23--27, 2001
\bibitem[1994]{Engels94}
 Engels, D., Cordis, L., \& K\"{o}hler, T.
 1994, Proc. IAU Symp. 161, ed. H. T. MacGillivray (Kluwer: Dordrecht), 317
\bibitem[1993]{Garcia93}
 Garcia, A. M., Bottinelli, L., Garnier, R., Gouguenheim, L., \& Paturel, G.
 1993, A\&AS, 97, 801
\bibitem[1999]{Grupe99}
 Grupe, D., Beuermann, K., Mannheim, K., \& Thomas  H.-C.
 1999, A\&A, 350, 895
\bibitem[1995]{Hagen95}
 Hagen, H.-J., Groote, D., Engels, D., \& Reimers, D.
 1995, A\&AS, 111, 195
\bibitem[2000]{Hopp00}
 Hopp, U., Engels, D., Green, R., et al.
 2000, A\&AS, 142, 417 ({\bf Paper~III})
\bibitem[1995]{Huchra95}
 Huchra, J. P., Geller, M. J., \& Corwin, H. G. Jr.
 1995, ApJS, 99, 391
\bibitem[1999]{Hunter99}
 Hunter, D., van Woerden, H., \& Gallagher, J.
 1999, AJ, 118, 2184
\bibitem[2001a]{Kniazev01}
 Kniazev, A. Y., Engels, D., Pustilnik, S. A. et al.
 2001a, A\&A, 366, 771 ({\bf Paper~IV})
\bibitem[2001b]{Kn-HS37}
 Kniazev, A. Y., Pustilnik, S. A., Ugryumov A.V. et al. 2001b,
 in preparation
\bibitem[1996]{KS96}
  Kobulnicky, H.A., Skillman, E.D. 1996, ApJ, 471, 211
\bibitem[1983]{Markarian83}
 Markarian, B. E., Lipovetsky, V. A., \& Stepanian, J. A.
 1983, Afz, 19, 29
\bibitem[1997]{McIntyre97}
 McIntyre, V. J.
 1997, Publ. Astron. Soc. Australia, 14, 122
\bibitem[1990]{Oke90}
 Oke, J. B.
 1990, AJ, 99, 1621
\bibitem[1995]{Pesch95}
 Pesch, P., Stephenson, C. B., \& MacConnell, D. J.
 1995, ApJS, 98, 41
\bibitem[2000]{Popescu00}
 Popescu, C. C., \& Hopp, U.
 2000, A\&AS, 142, 247
\bibitem[1996]{Popescu96}
 Popescu, C. C., Hopp, U., Hagen, H.-J., \& Els\"{a}sser, H.
 1996, A\&AS, 116, 43
\bibitem[1999]{Pustilnik99}
 Pustilnik, S. A., Engels, D., Ugryumov, A. V., et al.
 1999, A\&AS, 135, 299 ({\bf Paper~II})
\bibitem[2001]{Pustilnik01}
 Pustilnik, S. A., Kniazev, A. Y., Lipovetsky, V. A., \& Ugryumov, A. V.
 2001, A\&A, in press = astro-ph/0104334
\bibitem[1989]{Salzer89}
 Salzer, J. J., MacAlpine, G. M., \& Boroson, T. A.
 1989, ApJS, 70, 479
\bibitem[1995]{Salzer95}
 Salzer, J. J., Moody, J. W., Rosenberg, J. L., Gregory, S. A., \& Newberry, M. V.
 1995, AJ, 109, 2376
\bibitem[1994]{Stepanian94}
 Stepanian, J. A.
 1994, Proc. IAU Symp. 161, ed. H. T. MacGillivray (Kluwer: Dordrecht), 731
\bibitem[1997]{Thean97}
 Thean, A. H. C., Mundell, C. G., Pedlar, A., \& Nicholson, R. A.
 1997, MNRAS, 290, 15
\bibitem[1998]{Ugryumov98}
 Ugryumov, A. V., Pustilnik, S. A., Lipovetsky, V. A., Izotov, Yu. I.,
 \& Richter, G. M.
 1998, A\&AS, 131, 295
\bibitem[1999]{Ugryumov99}
 Ugryumov, A. V., Engels, D., Lipovetsky, V. A., et al.
 1999, A\&AS, 135, 511 ({\bf Paper~I})
\bibitem[2000]{Vennik2000}
 Vennik, J., Hopp, U., \& Popescu, C. C.
 2000, A\&AS, 142, 399

\end{thebibliography}
\end{document}